# An Epistemic Approach to Compositional Reasoning about Anonymity and Privacy

Yasuyuki Tsukada
NTT Communication Science
Laboratories, NTT Corporation
3-1 Morinosato-Wakamiya,
Atsugi, 243-0198 Japan
tsukada.yasuyuki@lab.ntt.co.jp

Hideki Sakurada
NTT Communication Science
Laboratories, NTT Corporation
3-1 Morinosato-Wakamiya,
Atsugi, 243-0198 Japan
sakurada.hideki@lab.ntt.co.jp

Ken Mano
NTT Communication Science
Laboratories, NTT Corporation
3-1 Morinosato-Wakamiya,
Atsugi, 243-0198 Japan
mano.ken@lab.ntt.co.jp

Yoshifumi Manabe
NTT Communication Science
Laboratories, NTT Corporation
2-4 Hikaridai, Seika-cho,
Kyoto, 619-0237 Japan
manabe.yoshifumi@lab.ntt.co.jp

## ABSTRACT

In this paper, we present an epistemic logic approach to the compositionality of several privacy-related information-hiding/disclosure properties. The properties considered here are anonymity, privacy, onymity, and identity. Our initial observation reveals that anonymity and privacy are not necessarily sequentially compositional; this means that even though a system comprising several sequential phases satisfies a certain unlinkability property in each phase, the entire system does not always enjoy a desired unlinkability property. We show that the compositionality can be guaranteed provided that the phases of the system satisfy what we call the independence assumptions. More specifically, we develop a series of theoretical case studies of what assumptions are sufficient to guarantee the sequential compositionality of various degrees of anonymity, privacy, onymity, and/or identity properties. Similar results for parallel composition are also discussed.

## Categories and Subject Descriptors

F.4.1 [**Mathematical Logic and Formal Languages**]: Mathematical Logic—*Modal logic*; D.2.4 [**Software Engineering**]: Software/Program Verification—*Formal methods*

## 1. INTRODUCTION

An information system generally consists of a number of subsystems. If some subsystems are shown to have certain formal properties and some others shown to have different properties, the question arises as to how we can deduce that the total system has certain formal properties. Or, more complicatedly, the system may possibly consist of a variety of subsystems that have various degrees of multiple properties. Thus, the concept of *compositionality* plays a key role in a modular approach to formal reasoning about complex information systems.

This paper deals with a logical approach to the compositionality of several privacy-related information-hiding/disclosure properties. Since privacy and related properties such as those discussed in [21, 2] have become crucial requirements for today's information systems, the compositionality of those properties has also become a concern. The properties considered here are *anonymity*, *privacy*, *onymity*, and *identity* (Fig. 1). Intuitively, we can understand anonymity to be the property of *hiding who* performed a certain specific action, privacy that of *hiding what* was performed by a certain specific agent, onymity that of *disclosing who* performed a certain specific action, and identity that of *disclosing what* was performed by a certain specific agent. A series of previous studies by Halpern and O'Neill [12], Mano *et al.* [19], and Tsukada *et al.* [26] showed that these properties can be formulated concisely in terms of *epistemic logic* (or *the modal logic of knowledge*) for multiagent systems.

For example, *sender anonymity* can be formulated in terms of our epistemic logic as

$$\theta(i, send(m)) \Rightarrow \bigwedge_{i' \in I_A} P_j[\theta(i', send(m))].$$

Here, $I_A$, called an *anonymity set*, denotes a set of possible senders. We read this formula as "if an agent $i$ sends a message $m$, then the observer $j$ thinks that it is possible that every agent $i'$ in $I_A$ performs the sending action." In other words, this formula means that the observer $j$ does not know who sends the message $m$. On the other hand, *message privacy* can be formulated as

$$\theta(i, send(m)) \Rightarrow \bigwedge_{a' \in A_I} P_j[\theta(i, a')].$$

Here, $A_I$, called a *privacy set*, denotes a set of possible sending actions, that is, $\{send(m') \mid m'$ is a possible message$\}$. This formula should be read as "if an agent $i$ sends a mes-sage $m$, then the observer $j$ thinks that it is possible that the agent $i$ performs every sending action $a'$ in $A_I$." In other words, this formula means that the observer $j$ does not know what message is sent from the agent $i$. We may say that these two properties—sender anonymity and message privacy—are "dual" because each of the above two formulas can be obtained from the other by interchanging "who" with "what," or more specifically, $I_A$ with $A_I$. We can also define onymity and identity as the "contrary" of anonymity and privacy, respectively, in terms of epistemic logic. Thus, epistemic logic enables us to succinctly describe formal specifications of various privacy-related information-hiding/disclosure properties of information systems.

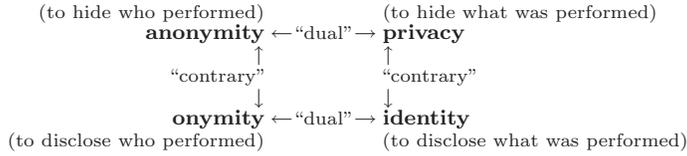

Figure 1: Privacy-related information-hiding/disclosure properties.

In this paper, the epistemic logic approach developed in [12, 19, 26] is further exploited to discuss the compositionality of multiple properties comprising anonymity, privacy, onymity, and identity. More specifically, the contributions of this paper can be summarized as follows. First, we indicate that anonymity and privacy are not necessarily sequentially compositional. (This may be contrary to our intuition, because we might think that anonymity/privacy can be reinforced by sequentially connecting anonymous/private communication channels.) To show this indication, we introduce, as a motivating example, an abstract model of an anonymous members-only bulletin board system, which comprises two sequential phases, namely, the registration and posting phases. We show that the composition of anonymity in the registration phase and privacy in the posting phase does not necessarily induce anonymity or privacy in the entire system. If we regard anonymity and privacy as special cases of *unlinkability*, this indication can be paraphrased by saying that even though a system comprising several sequential phases satisfies a certain unlinkability property in each phase, the system as a whole does not always enjoy a desired unlinkability property. For example, our epistemic logic approach shows that a chain $M_1 * M_2$ of two *mix-servers* [5] does not necessarily guarantee unlinkability between incoming and outgoing messages even though both $M_1$ and $M_2$ do. This non-compositionality of unlinkability can be viewed as being analogous to the non-transitivity of inequality: $a \neq b$ and $b \neq c$ do not necessarily imply $a \neq c$. Second, we show that the sequential compositionality of anonymity and privacy can be guaranteed provided that the phases of the system satisfy what we call the *independence assumptions*. We develop a series of case studies of what assumptions are sufficient to guarantee the sequential compositionality of various degrees of anonymity, privacy, onymity, and/or identity properties. These compositionality results are summarized in Table 1. Third, we show that similar compositionality results can be obtained for parallel composition. We demonstrate that some variations of independence assumptions also play important roles in guaranteeing the parallel compositionality of anonymity and privacy.

## Related Work

A considerable amount of substantial research on the measurement, characterization, and taxonomy of privacy and related information-hiding/disclosure properties has been undertaken from various standpoints [7, 23, 8, 25, 14, 17, 21, 30]. The present paper focuses on formal approaches to privacy-related properties, since our primary motivation is to contribute to the development of a new methodology for the formal verification of these properties.

Formal approaches to privacy-related information-hiding properties go back to the seminal work of Schneider and Sidiropoulos [22], who formulated the concept of *strong anonymity* in terms of a process calculus called CSP. Since then, this concept has been further developed and elaborated in various computational or logical frameworks such as ACP [20], applied $\pi$ calculus [6], I/O-automata [16], category theory [13], and epistemic logic [24, 27, 10, 15, 29, 1, 28, 18, 4, 3].

Although the approach presented in this paper shares a common style of anonymity definitions with these epistemic logic approaches, it directly builds on the approach described by Halpern and O'Neill [12]. Within Halpern and O'Neill's framework, Mano *et al.* [19] formulated privacy as the dual of anonymity and showed that these two properties can be related by a newly proposed information-hiding property called *role interchangeability*. They proved the role-interchangeability property of a practical electronic voting protocol, thereby demonstrating the *voter anonymity* and *vote privacy* properties of the protocol. Further, Tsukada *et al.* [26] considered the logical contraries of anonymity and privacy, thereby giving formal definitions of onymity and identity. In particular, they showed that some weak forms of anonymity and privacy are compatible with some weak forms of onymity and identity, respectively. They also discussed the relationships between their proposed definitions and existing standard terminology, in particular Pfitzmann and Hansen's consolidated proposal [21]. The epistemic logic approach developed in [12, 19, 26] has recently been extended by Goriac [11], where a wider spectrum of privacy-related properties including *undetectability*, *unobservability*, and *pseudonymity* are formulated and discussed.

## 2. EPISTEMIC DEFINITIONS OF ANONYMITY AND PRIVACY

We briefly review epistemic logic for multiagent systems. Notions and terminologies are borrowed from [9, 12].

A *multiagent system* consists of $n$ agents with their *local states* and develops over time. We assume that an agent's local state encapsulates all the information to which the agent has access. Let $I = \{i_1, \ldots, i_n\}$ be the set of $n$ agents. A *global state* is defined as the tuple $(s_{i_1}, \ldots, s_{i_n})$ with all local states from $i_1$ to $i_n$. A *run* is a function from *time*, ranging over the natural numbers, to global states. A *point* is a pair $(r, m)$ comprising a run $r$ and a time $m$, and the global state at a point $(r, m)$ is denoted by $r(m)$. The function $r_x$ of $m$ is the projection of $r(m)$ to $x$'s component, so that $r_x(m) = s_x$ if $r(m) = (s_{i_1}, \ldots, s_{i_n})$ for $x = i_1, \ldots, i_n$. A *system* is a set of runs. The set of all points in a system $\mathcal{R}$ is denoted by $\mathcal{P}(\mathcal{R})$.

In a multiagent system, we can define the knowledge of an agent on the basis of the indistinguishability of the state for the agent. Given a system $\mathcal{R}$ and an agent $i$, let $\mathcal{K}_i(r, m)$ be the set of points in $\mathcal{P}(\mathcal{R})$ that $i$ thinks are possible at $(r, m)$; that is, $\mathcal{K}_i(r, m) = \{(r', m') \in \mathcal{P}(\mathcal{R}) \mid (r', m') \sim_i (r, m)\}$, where $(r', m') \sim_i (r, m)$ means that $r'_i(m') = r_i(m)$. We can say that an agent $i$ "knows" $\phi$ at a point $(r, m)$ if $\phi$ is true at all points in $\mathcal{K}_i(r, m)$.

The *formulas* of epistemic logic are inductively constructed from a set $\Phi$ of *primitive propositions* (such as "the key is $k$" or "an agent $i$ sent a message $m$ to an agent $j$"), the usual logical connectives, and an epistemic operator $K_i$ that represents the knowledge of agent $i$. The meaning of each formula can be determined when each primitive proposition is given an interpretation. An *interpreted system* $\mathcal{I}$ consists of a pair $(\mathcal{R}, \pi)$ comprising a system $\mathcal{R}$ and an *interpretation* $\pi$ that maps each point to the truth-value assignment function for $\Phi$ for the point. In other words, $(\pi(r, m))(p) \in \{true, false\}$ for each $p \in \Phi$ and $(r, m) \in \mathcal{P}(\mathcal{R})$. Given an interpreted system $\mathcal{I} = (\mathcal{R}, \pi)$ and a point $(r, m)$ in $\mathcal{R}$, we define what it means for a formula $\phi$ to be true at $(r, m)$ in $\mathcal{I}$ by induction on the structure of formulas. Typical cases are as follows: $(\mathcal{I}, r, m) \models p$ if $(\pi(r, m))(p) = true$; $(\mathcal{I}, r, m) \models \neg \phi$ if $(\mathcal{I}, r, m) \not\models \phi$; $(\mathcal{I}, r, m) \models \phi \wedge \psi$ if $(\mathcal{I}, r, m) \models \phi$ and $(\mathcal{I}, r, m) \models \psi$; $(\mathcal{I}, r, m) \models K_i \phi$ if $(\mathcal{I}, r', m') \models \phi$ for all $(r', m') \in \mathcal{K}_i(r, m)$. In addition to $K_i \phi$, which means that $i$ knows $\phi$, we also use $P_i \phi$ as an abbreviation of $\neg K_i \neg \phi$, which means that $i$ thinks that $\phi$ is possible. We also write $\mathcal{I} \models \phi$ if $(\mathcal{I}, r, m) \models \phi$ holds for every point $(r, m)$ in $\mathcal{I}$.

In the rest of the paper, we consider that the set $A$ of *actions* is also associated with each system. We assume that $i, i', j, j', \ldots$ range over agents while $a, a', b, b', \ldots$ range over actions. Following [12], we use a primitive proposition of the form $\theta(i, a)$, which denotes that "an agent $i$ has performed an action $a$, or will perform $a$ in the future." Note that the truth value of $\theta(i, a)$ depends on the run, but not on the time; that is, if $(\mathcal{I}, r, m) \models \theta(i, a)$ holds for some $m$, then $(\mathcal{I}, r, m') \models \theta(i, a)$ also holds for every $m'$.

Below we review the formal definitions of anonymity, privacy, onymity, and identity in terms of epistemic logic for multiagent systems. For full details, see [12, 19, 26].

### Anonymity

We say that an action $a$ performed by an agent $i$ is *anonymous up to* an *anonymity set* $I_A \subseteq I$ with respect to an agent $j$ in the interpreted system $\mathcal{I}$ if $\mathcal{I} \models \theta(i, a) \Rightarrow \bigwedge_{i' \in I_A} P_j[\theta(i', a)]$ holds. Intuitively, anonymity up to $I_A$ means that, from $j$'s viewpoint, $a$ could have been performed by anybody in $I_A$. A typical example of anonymity of this form is *sender anonymity*, which is explained in Sect. 1.

We also say that an action $a$ performed by an agent $i$ is *minimally anonymous* with respect to an agent $j$ in the interpreted system $\mathcal{I}$ if $\mathcal{I} \models \theta(i, a) \Rightarrow P_j[\neg \theta(i, a)]$ holds. Intuitively, minimal anonymity means that, from $j$'s viewpoint, $a$ could not have been performed by $i$. Consider that our built-in proposition $\theta(i, a)$ expresses a specific form of "link" between an agent $i$ and an action $a$. Then, we can observe that minimal anonymity is very close to a specific form of the "unlinkability" property that was stipulated by Pfitzmann and Hansen [21]. This observation was elaborated in [26].

### Privacy

Privacy properties can be obtained from anonymity properties by applying the operation of taking the agent/action reversal dual, that is, the operation that replaces a set of agents with a set of actions. For example, we say that an agent $i$ performing an action $a$ is *private up to* a *privacy set* $A_I \subseteq A$ with respect to an agent $j$ in the interpreted system $\mathcal{I}$ if $\mathcal{I} \models \theta(i, a) \Rightarrow \bigwedge_{a' \in A_I} P_j[\theta(i, a')]$ holds. Intuitively, privacy up to $A_I$ means that, from $j$'s viewpoint, $i$ could have performed any action in $A_I$. A typical example is *message privacy*, which is explained in Sect. 1.

We also say that an agent $i$ performing an action $a$ is *minimally private* with respect to an agent $j$ in the interpreted system $\mathcal{I}$ if $\mathcal{I} \models \theta(i, a) \Rightarrow P_j[\neg \theta(i, a)]$ holds. Note that minimal privacy is equivalent to its dual, that is, minimal anonymity.

### Role Interchangeability

Role interchangeability means that, as far as an agent $j$ is concerned, two agents $i$ and $i'$ could interchange their roles, that is, the actions they performed. Specifically, a pair $(i, a)$ comprising an agent $i$ and an action $a$ is *role interchangeable* with respect to an agent $j$ in the interpreted system $\mathcal{I}$ if $\mathcal{I} \models \theta(i, a) \Rightarrow \bigwedge_{i' \in I/\{j\}} \bigwedge_{a' \in A} (\theta(i', a') \Rightarrow P_j[\theta(i', a) \wedge \theta(i, a')])$ holds. Despite the similarity between role interchangeability and anonymity/privacy, they are not equiexpressive. We can prove that role interchangeability implies both anonymity and privacy under some appropriate conditions [19].

### Onymity

By the "contrary" of a formula of the form $\theta(i, a) \Rightarrow \Gamma$, we mean the formula $\theta(i, a) \Rightarrow \neg \Gamma$. By taking the contrary of the formulas defining anonymity, we can obtain definitions of onymity. We only show below the contrary of minimal anonymity. We say that an action $a$ performed by an agent $i$ is *maximally onymous* with respect to an agent $j$ in the interpreted system $\mathcal{I}$ if $\mathcal{I} \models \theta(i, a) \Rightarrow K_j[\theta(i, a)]$ holds. Intuitively, maximal onymity means that $j$ knows that $i$ has performed $a$. This definition corresponds to our observation that onymity generally means that the agent who performs the action is disclosed. We can see that onymity is closely related to personal authentication.

### Identity

Identity properties, which are closely related to attribute authentication, can be obtained as the contrary of privacy properties or as the dual of onymity properties. Below we only show the contrary of minimal privacy. We say that an agent $i$ performing an action $a$ is *maximally identified* with respect to an agent $j$ in the interpreted system $\mathcal{I}$ if $\mathcal{I} \models \theta(i, a) \Rightarrow K_j[\theta(i, a)]$ holds. Note that maximal identity is equivalent to its dual, that is, maximal onymity.

The definitions of the properties presented above and their known relationships are summarized in Fig. 2. For example, role interchangeability implies anonymity up to $I_A$, which also implies minimal anonymity. Note that every implication

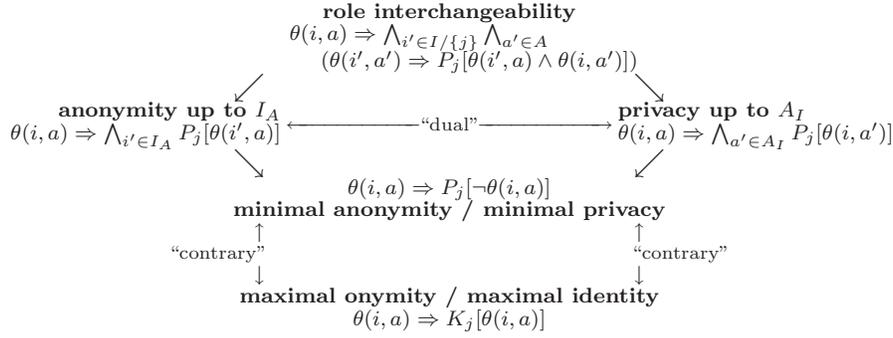

Figure 2: Formal definitions of some privacy-related information-hiding/disclosure properties.

described here is conditional. A more detailed version of this figure can be found in [26].

## 3. SEQUENTIAL COMPOSITIONALITY OF ANONYMITY AND PRIVACY

As a motivating example for discussion of sequential compositionality, consider an abstract model of an anonymous members-only bulletin board system (Fig. 3). Suppose that the set of agents includes two disjoint subsets $I_R$ and $I_P$ of *real names* and *pseudonyms*, respectively. Each real-name agent can register several pseudonyms to use; the correspondence between real names and pseudonyms is expressed by using $\theta(i, use(k))$, which means that a real $i$ can use a pseudonym $k$. Besides $I_R$ and $I_P$, we also introduce the domain $C$ of possible articles. Each real-name agent uses some of its pseudonyms and posts some articles to a bulletin board. We express this as $\theta(k, post(c))$, which means that a pseudonym $k$ posts an article $c$. When a real-name agent $i$ uses a pseudonym $k$ and $k$ posts an article $c$, we say that $i$ submits $c$. This is formulated as $\mathcal{I} \models \theta(i, submit(c)) \Leftrightarrow \bigvee_{k \in I_P}(\theta(i, use(k)) \wedge \theta(k, post(c)))$. Two sets $\{post(c) \mid c \in C\}$ and $\{submit(c) \mid c \in C\}$ of actions are denoted by $A_P$ and $A_S$, respectively.

Although this is initially given as a model of an anonymous bulletin board system, it is quite abstract and can serve as a model for a more general class of systems, provided that it is appropriately modified. For example, if $\theta(i, use(k))$ is interpreted as meaning that a voter $i$ is authorized to use a pseudonym $k$ for voting and $\theta(k, post(c))$ is interpreted as meaning that $k$ casts a ballot $c$ for some candidate, then this will be regarded as a model of a voting system. (Of course, some appropriate assumptions will be required. For example, to guarantee eligibility, we must assume that each voter uses at most one pseudonym and each pseudonym also casts at most one ballot.) Furthermore, if $\theta(i, use(k))$ is interpreted as meaning that the first mix-server takes an incoming message $i$ and produces an outgoing message $k$ and if $\theta(k, post(c))$ is interpreted as meaning that the second mix-server takes an incoming message $k$ and produces an outgoing message $c$, then this will be regarded as a model of a chain of two mix-servers.

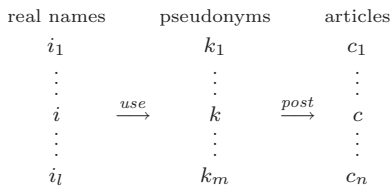

Figure 3: An anonymous members-only bulletin board system.

We shall consider several typical cases where different combinations of privacy-related properties are owned by each registration and posting phase (Table 1). Below we concentrate on some main specific cases (Cases 1 to 5). The other cases are discussed in Appendix A. Intuitively, when registration is anonymous and posting is private (Case 1), the entire system appears to have good anonymity/privacy properties. However, this conjecture is refuted. Indeed, assume that an observer has some presupposed background knowledge that a real-name agent $i$ will never submit an improper article $c$. Then, even though the observer thinks that any real-name agents including $i$ could have used a pseudonym $k$ and that $k$ could have posted any articles including $c$, the observer never thinks that $i$ could have submitted $c$. More formally, the following holds.

CLAIM 3.1. *There is an interpreted system that satisfies the following: (1) every action $use(k)$ performed by $i$ is anonymous up to $I_R$ with respect to an observer $j$; (2) every agent $k$ performing $post(c)$ is private up to $A_P$ with respect to $j$; (3) some action $submit(c)$ performed by $i$ is not anonymous up to $I_R$; (4) some agent $i$ performing $submit(c)$ is not private up to $A_S$.*

PROOF. Suppose that $I_R = \{i_1, i_2\}$, $I_P = \{k_1, k_2\}$, $A_P = \{post(c_1), post(c_2)\}$, and $A_S = \{submit(c_1), submit(c_2)\}$. Consider an interpreted system consisting of two runs $r_1$ and $r_2$. In $r_1$, the following are true: $\theta(i_1, use(k_1))$, $\theta(k_1, post(c_1))$, $\theta(i_2, use(k_2))$, and $\theta(k_2, post(c_2))$. In $r_2$, the following are true: $\theta(i_1, use(k_2))$, $\theta(k_2, post(c_1))$, $\theta(i_2, use(k_1))$, and $\theta(k_1, post(c_2))$. We also assume that the two runs are indistinguishable from the observer $j$'s viewpoint, that is, more precisely, $(r_1, m) \sim_j (r_2, m)$ holds for each $m$. Then, it is immediately seen that (1) and (2) hold. Furthermore, (3) and (4) also hold because $\theta(i_1, submit(c_2))$ is neither true in $r_1$ nor true in $r_2$ and because $\theta(i_2, submit(c_1))$ is neither true in $r_1$ nor true in $r_2$. In other words, the observer can have "presupposed background knowledge" that $i_1$ never submits $c_2$, and $i_2$ never submits $c_1$. □

*Remark 1.* The observations above, in particular, the construction of $\{r_1, r_2\}$ shown in the proof of Claim 3.1, can

242

Table 1: Sequential Compositionality: Twelve Cases

| | | Assumption | Registration | Posting | Total |
|---|---|---|---|---|---|
| Case 1 | (Claim 3.1) | — | Anonymous up to $I_R$ | Private up to $A_P$ | — |
| Case 2 | (Claim 3.2) | Independent | — | Private up to $A_P$ | Private up to $A_S$ |
| Case 3 | (Claim 3.3) | Independent | Anonymous up to $I_R$ | — | Anonymous up to $I_R$ |
| Case 4 | (Claim 3.4) | — | Maximally onymous | Private up to $A_P$ | Private up to $A_S$ |
| Case 5 | (Claim 3.5) | — | Anonymous up to $I_R$ | Maximally identified | Anonymous up to $I_R$ |
| Case 6 | (Claim A.1) | Pairwise independent | — | Role interchangeable | Role interchangeable |
| Case 7 | (Claim A.2) | Pairwise independent | Role interchangeable | — | Role interchangeable |
| Case 8 | (Claim A.3) | Independent & Exhaustive posting & Exclusive $i$ and $post(c)$ | — | Minimally private | Minimally private |
| Case 9 | (Claim A.4) | Independent & Exhaustive registration & Exclusive $i$ and $post(c)$ | Minimally anonymous | — | Minimally anonymous |
| Case 10 | (Claim A.5) | Exhaustive posting & Exclusive $i$ and $post(c)$ | Maximally onymous | Minimally private | Minimally private |
| Case 11 | (Claim A.6) | Exhaustive registration & Exclusive $i$ and $post(c)$ | Minimally anonymous | Maximally identified | Minimally anonymous |
| Case 12 | (Claim A.7) | — | Maximally onymous | Maximally identified | Maximally onymous/identified |

be extended to consider other examples where anonymity/privacy properties are not sequentially compositional. For example, we can say that a chain $M_1 * M_2$ of two mix-servers does not necessarily guarantee unlinkability between incoming and outgoing messages even though $M_1$ and $M_2$ do individually. Indeed, if $M_2$ is the "inverse" $M_1^{-1}$ of $M_1$, then $M_1 * M_1^{-1}$ becomes an identity and thus provides obvious linkability, even though both $M_1$ and $M_1^{-1}$ guarantee unlinkability.

On the basis of the above discussion, we introduce "independence" assumptions so that anonymity/privacy in the entire system can be obtained quite directly from anonymity/privacy in the registration/posting phases. The registration and posting phases in an anonymous bulletin board system $\mathcal{I}$ are *independent* with respect to an observer $j$ if

$$\mathcal{I} \models P_j[\theta(i, use(k))] \wedge P_j[\theta(k', post(c))]$$
$$\Rightarrow P_j[\theta(i, use(k)) \wedge \theta(k', post(c))]$$

holds for every $i$, $k$, $k'$, and $c$. This is analogous to the independence of two events in probability theory: two events $A$ and $B$ are independent if $\Pr(A)\Pr(B) = \Pr(A \cap B)$. The independence assumption can be regarded as meaning that the observer has no specific "presupposed background knowledge."

*Example 1.* In the system $\{r_1, r_2\}$ shown in the proof of Claim 3.1, the registration and posting phases are not independent. To guarantee independence, we can extend the system so that it has four indistinguishable runs $\{r_1, r_2, r_3, r_4\}$ (Fig. 4). In $r_3$, the following are true: $\theta(i_1, use(k_1))$, $\theta(k_1, post(c_2))$, $\theta(i_2, use(k_2))$, and $\theta(k_2, post(c_1))$. In $r_4$, the following are true: $\theta(i_1, use(k_2))$, $\theta(k_2, post(c_2))$, $\theta(i_2, use(k_1))$, and $\theta(k_1, post(c_1))$. Alternatively, we can also obtain a system $\{r_1, r_2, r_5, r_6, r_7, r_8\}$ of indistinguishable runs that has the independence property. Similarly, a system $\{r_1, r_2, r_9, r_{10}, r_{11}, r_{12}\}$ of indistinguishable runs also has the independence property.

We also discuss, in Appendix C, that independence could be viewed by itself as a "meta-level" abstraction of anonymity or privacy.

The following two lemmas are "dual" and show some obvious sufficient conditions for independence. Hereafter, the proofs of the "dual" of proved lemmas or claims are omitted, since they can be straightforwardly obtained from the original proofs via duality.

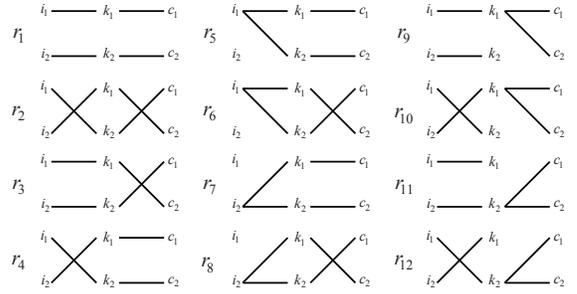

**Figure 4:** Systems $\{r_1, r_2, r_3, r_4\}$, $\{r_1, r_2, r_5, r_6, r_7, r_8\}$, and $\{r_1, r_2, r_9, r_{10}, r_{11}, r_{12}\}$ **of runs satisfy the independence property.**

LEMMA 3.1. *If every action $use(k)$ performed by $i$ is maximally onymous with respect to an observer $j$, the registration and posting phases are independent with respect to $j$.*

PROOF. Suppose that $(\mathcal{I}, r, m) \models P_j[\theta(i, use(k))] \wedge P_j[\theta(k', post(c))]$. Then, $\theta(i, use(k))$ holds at some point $(r', m')$ such that $(r', m') \sim_j (r, m)$, and $\theta(k', post(c))$ also holds at some point $(r'', m'')$ such that $(r'', m'') \sim_j (r, m)$. Since $use(k)$ performed by $i$ is maximally onymous and $\theta(i, use(k))$ holds at $(r', m')$, $\theta(i, use(k))$ also holds at $(r'', m'')$. In other words, $(\mathcal{I}, r'', m'') \models \theta(i, use(k)) \wedge \theta(k', post(c))$ holds. Thus, we have proved that $(\mathcal{I}, r, m) \models P_j[\theta(i, use(k)) \wedge \theta(k', post(c))]$. □

LEMMA 3.2. *If every agent $k$ performing $post(c)$ is maximally identified with respect to an observer $j$, the registration and posting phases are independent with respect to $j$.*

Case 2 in Table 1 indicates that if the posting phase guarantees privacy, then so does the entire system, provided that the posting and registration phases are independent.

CLAIM 3.2. *Assume that the registration and posting phases are independent with respect to an observer $j$. Also*

suppose that every agent $k$ performing $post(c)$ is private up to $A_P$ with respect to $j$. Then, every agent $i$ performing $submit(c)$ is private up to $A_S$.

PROOF. Suppose that $(\mathcal{I}, r, m) \models \theta(i, submit(c))$. Then, there exists some $k$ in $I_P$ such that $(\mathcal{I}, r, m) \models \theta(i, use(k)) \wedge \theta(k, post(c))$. From $(\mathcal{I}, r, m) \models \theta(i, use(k))$, it is immediate to see that $(\mathcal{I}, r, m) \models P_j[\theta(i, use(k))]$. Because every $k$ performing $post(c)$ is private up to $A_P$ and because $(\mathcal{I}, r, m) \models \theta(k, post(c))$, we can say that for every possible article $c'$, $(\mathcal{I}, r, m) \models P_j[\theta(k, post(c'))]$ holds. So, by virtue of the independence assumption, $(\mathcal{I}, r, m) \models P_j[\theta(i, use(k)) \wedge \theta(k, post(c'))]$ holds. That is, $(\mathcal{I}, r, m) \models P_j[\theta(i, submit(c'))]$ holds. Since $c'$ is arbitrary, we have proved that $(\mathcal{I}, r, m) \models \bigwedge_{a' \in A_S} P_j[\theta(i, a')]$. □

Case 3 in Table 1 is a "dual" of Case 2. It means that if the registration phase guarantees anonymity, then so does the entire system, provided that the posting and registration phases are independent.

CLAIM 3.3. *Assume that the registration and posting phases are independent with respect to an observer $j$. Also suppose that every action $use(k)$ performed by $i$ is anonymous up to $I_R$ with respect to $j$. Then, every action $submit(c)$ performed by $i$ is anonymous up to $I_R$.*

In the view of Lemma 3.1, Case 4 can be regarded as a special case of Case 2. More specifically, the following claim directly follows from Lemma 3.1 and Claim 3.2. It indicates that if the posting phase guarantees privacy, then so does the entire system, even though each registered pseudonym is linked to the corresponding real name.

CLAIM 3.4. *Suppose that every action $use(k)$ performed by $i$ is maximally onymous with respect to an observer $j$. Also suppose that every agent $k$ performing $post(c)$ is private up to $A_P$ with respect to $j$. Then, every agent $i$ performing $submit(c)$ is private up to $A_S$.*

Case 5 is a "dual" of Case 4. It can also be regarded, in the view of Lemma 3.2, as a special case of Case 3. It means that if the registration phase guarantees anonymity, then so does the entire system, even though each article is linked to the pseudonym who posted it.

CLAIM 3.5. *Suppose that every action $use(k)$ performed by $i$ is anonymous up to $I_R$ with respect to an observer $j$. Also suppose that every agent $k$ performing $post(c)$ is maximally identified with respect to $j$. Then, every action $submit(c)$ performed by $i$ is anonymous up to $I_R$.*

## 4. PARALLEL COMPOSITIONALITY OF ANONYMITY AND PRIVACY

By the *parallel composition* of $act_a(c)$ performed by $i$ and $act_b(c)$ performed by $i$, we generally mean the action $act_p(c)$ performed by $i$ that is introduced by $\theta(i, act_p(c)) \Leftrightarrow \theta(i, act_a(c)) \wedge \theta(i, act_b(c))$. We denote three sets $\{act_a(c) \mid c\}$, $\{act_b(c) \mid c\}$, and $\{act_p(c) \mid c\}$ of actions by $A_a$, $A_b$, and $A_p$, respectively.

*Example 2.* Consider the following situation. A special prosecution team has pursued their probe into the hideout of a radical and has found out a time bomb $c$ that seems to have been provided by a sympathizer $i$. The urgent mission of the team is to determine $i$ performing an action $give(c)$. The essential parts of the bomb $c$ are a timer and gunpowder. The sympathizer seems to have bought the timer and have synthesized the gunpowder, thereby producing the time bomb. Thus, the following definition is obtained: $\theta(i, give(c)) \Leftrightarrow \theta(i, buy\_timer(c)) \wedge \theta(i, synthesize\_gunpowder(c))$. A concern here is how some (an)onymity property of $give(c)$ can be deduced from the (an)onymity properties of $buy\_timer(c)$ and $synthesize\_gunpowder(c)$.

Table 2 shows some cases where different combinations of privacy-related properties are owned by $act_a$ and $act_b$. As for the case of sequential composition, the parallel compositionality of anonymity or privacy does not generally hold without some appropriate forms of independence assumptions. We say that $act_a$ and $act_b$ are *independent* with respect to an observer $j$ in a system $\mathcal{I}$ if $\mathcal{I} \models P_j[\theta(i, act_a(c))] \wedge P_j[\theta(i, act_b(c))] \Rightarrow P_j[\theta(i, act_a(c)) \wedge \theta(i, act_b(c))]$ holds for every $i$ and $c$. Roughly speaking, the independence means that $act_a$ and $act_b$ are not exclusive. Below we show that the independence assumption plays an essential role in Case I and its dual, Case II. The other cases are discussed in Appendix B.

CLAIM 4.1. *Assume that $act_a$ and $act_b$ are independent with respect to an observer $j$. Also suppose that $i$ performing $act_a(c)$ is private up to $A_a$ with respect to $j$ and $i$ performing $act_b(c)$ is private up to $A_b$ with respect to $j$. Then, $i$ performing $act_p(c)$ is private up to $A_p$ with respect to $j$.*

PROOF. Suppose that $(\mathcal{I}, r, m) \models \theta(i, act_p(c))$. Then, $(\mathcal{I}, r, m) \models \theta(i, act_a(c)) \wedge \theta(i, act_b(c))$ holds. By the assumption of privacy, we have $(\mathcal{I}, r, m) \models P_j[\theta(i, act_a(c'))] \wedge P_j[\theta(i, act_b(c'))]$ for every $c'$. By the independence assumption, $(\mathcal{I}, r, m) \models P_j[\theta(i, act_a(c')) \wedge \theta(i, act_b(c'))]$, that is, $(\mathcal{I}, r, m) \models P_j[\theta(i, act_p(c'))]$ holds. Since $c'$ is arbitrary, we have proved the claim. □

CLAIM 4.2. *Assume that $act_a$ and $act_b$ are independent with respect to an observer $j$. Also suppose that $act_a(c)$ performed by $i$ and $act_b(c)$ performed by $i$ are anonymous up to $I_a$ and $I_b$, respectively. Then, $act_p(c)$ performed by $i$ is anonymous up to $I_a \cap I_b$ with respect to $j$.*

*Example 3.* Consider the situation described in Example 2. Claim 4.2 indicates that $give(c)$ can be onymous even though both $buy\_timer(c)$ and $synthesize\_gunpowder(c)$ are anonymous. This can happen when $buy\_timer$ and $synthesize\_gunpowder$ are not independent, that is, when some suspect is considered to be unable to perform both actions for some reason.

## 5. CONCLUSION

Building on an epistemic-logic formalism, we have discussed the compositionality of several privacy-related information-hiding/disclosure properties. We have pointed out that anonymity and privacy are not necessarily sequentially compositional and have indicated that the independence assumptions can guarantee the compositionality. We have also developed a series of theoretical case studies on the conditions that are sufficient to guarantee the sequential compositionality of various degrees of anonymity, privacy, onymity, and/or identity. Similar compositionality results have also been shown for parallel composition.

Table 2: Parallel Compositionality: Five Cases

|  |  | Assumption | $act_a$ | $act_b$ | $act_p$ (Total) |
|---|---|---|---|---|---|
| Case I | (Claim 4.1) | Independent | Private up to $A_a$ | Private up to $A_b$ | Private up to $A_p$ |
| Case II | (Claim 4.2) | Independent | Anonymous up to $I_a$ | Anonymous up to $I_b$ | Anonymous up to $I_a \cap I_b$ |
| Case III | (Claim B.1) | — | — | Minimally anonymous/private | Minimally anonymous/private |
| Case IV | (Claim B.1) | — | Minimally anonymous/private | — | Minimally anonymous/private |
| Case V | (Claim B.2) | — | Maximally onymous/identified | Maximally onymous/identified | Maximally onymous/identified |

Future work will include a discussion of compositionality in terms of the probabilistic extension [12] of epistemic logic. To substantiate the practical value of our approach, a detailed analysis of real world examples should be carried out.

# APPENDIX

## A. SEQUENTIAL COMPOSITIONALITY: MORE CASES

In this appendix, we discuss Cases 6 to 12 shown in Table 1.

We first introduce some additional conditions regarding our motivating example of an anonymous members-only bulletin board system. We say that an action $post(c)$ is *exclusive* if $post(c)$ is performed by at most one pseudonym in each run, that is, $\mathcal{I} \models \bigwedge_{k \neq k'} \neg[\theta(k, post(c)) \wedge \theta(k', post(c))]$ holds. For example, if we consider that each article $c$ is labeled and identified with an article ID number, we will accordingly assume that each $post(c)$ is exclusive. We also say that an action $use(k)$ is *exclusive* if $use(k)$ is performed by at most one real-name agent in each run. For example, if we want to avoid the use of bogus pseudonyms, we will assume that each $use(k)$ is exclusive. Similarly, we say that a real-name agent $i$ is *exclusive* if $i$ performs at most one $use(k)$ action in each run, that is, $\mathcal{I} \models \bigwedge_{k \neq k'} \neg[\theta(i, use(k)) \wedge \theta(i, use(k'))]$ holds. We also say that a pseudonym $k$ is *exclusive* if $k$ performs at most one $post(c)$ action in each run.

We also say that the posting phase is *exhaustive* provided that every article $c \in C$ has been posted by some pseudonyms. This is formulated as $\mathcal{I} \models \bigwedge_{c \in C} \bigvee_{k \in I_P} \theta(k, post(c))$. Similarly, we say that the registration phase is *exhaustive* provided that every real-name agent $i \in I_R$ uses some pseudonyms. This is formulated as $\mathcal{I} \models \bigwedge_{i \in I_R} \bigvee_{k \in I_P} \theta(i, use(k))$.

We also extend the independence assumption so as to deal with Cases 6 to 11. First, the independence assumption can immediately be extended to a disjunctive form.

LEMMA A.1. *If the registration and posting phases in $\mathcal{I}$ are independent with respect to an observer $j$, then the following holds for arbitrary $i_p$, $k_p$, $k'_q$, and $c_q$:*

$$\mathcal{I} \models P_j[\bigvee_p \theta(i_p, use(k_p))] \wedge P_j[\bigvee_q \theta(k'_q, post(c_q))]$$
$$\Rightarrow P_j[(\bigvee_p \theta(i_p, use(k_p))) \wedge (\bigvee_q \theta(k'_q, post(c_q)))].$$

PROOF. Suppose that $(\mathcal{I}, r, m) \models P_j[\vee_p \theta(i_p, use(k_p))]$ and $(\mathcal{I}, r, m) \models P_j[\vee_q \theta(k'_q, post(c_q))]$. This means that there exist some point $(r', m')$ and $p$ such that $\theta(i_p, use(k_p))$ holds at $(r', m')$ and $(r', m') \sim_j (r, m)$. Further, there exist some point $(r'', m'')$ and $q$ such that $\theta(k'_q, post(c_q))$ holds at $(r'', m'')$ and $(r'', m'') \sim_j (r, m)$. Then, by the independence assumption, there exists some point $(r''', m''')$ such that $\theta(i_p, use(k_p)) \wedge \theta(k'_q, post(c_q))$ holds at $(r''', m''')$ and $(r''', m''') \sim_j (r, m)$. This concludes the proof. □

Further, the independence assumption can be extended to "positive-negative" and "negative-positive" forms.

LEMMA A.2. *Assume that the registration and posting phases in $\mathcal{I}$ are independent with respect to an observer $j$. Also assume that the posting phase is exhaustive and that every posting action $post(c)$ is exclusive. Then, $\mathcal{I} \models P_j[\theta(i, use(k))] \wedge P_j[\neg \theta(k', post(c))] \Rightarrow P_j[\theta(i, use(k)) \wedge \neg \theta(k', post(c))]$ holds for every $i$, $k$, $k'$, and $c$.*

PROOF. Since the posting phase is exhaustive, every $c$ must have been posted by some pseudonyms in each run. Further, since $post(c)$ is exclusive, a uniquely determined pseudonym must have posted it in each run. In other words, $\neg\theta(k', post(c))$ can be equivalently expressed as a formula of the form $\vee_{k'_q \neq k'} \theta(k'_q, post(c))$. Hence, the lemma immediately follows from Lemma A.1. □

LEMMA A.3. *Assume that the registration and posting phases in $\mathcal{I}$ are independent with respect to an observer $j$. Also assume that the registration phase is exhaustive and that every real-name agent $i$ is exclusive. Then, $\mathcal{I} \models P_j[\neg \theta(i, use(k))] \wedge P_j[\theta(k', post(c))] \Rightarrow P_j[\neg \theta(i, use(k)) \wedge \theta(k', post(c))]$ holds for every $i$, $k$, $k'$, and $c$.*

In some cases, we require a stronger form of the independence assumption to prove compositionality results. Indeed, we need the binarily conjunctive form of the assumption. More specifically, the registration and posting phases in an anonymous bulletin board system $\mathcal{I}$ are *pairwise independent* with respect to an observer $j$ if

$$\mathcal{I} \models P_j[\bigwedge_{m \in \{0,1\}} \theta(i_m, use(k_m))] \wedge P_j[\bigwedge_{n \in \{0,1\}} \theta(k'_n, post(c_n))]$$
$$\Rightarrow P_j[(\bigwedge_{m \in \{0,1\}} \theta(i_m, use(k_m))) \wedge (\bigwedge_{n \in \{0,1\}} \theta(k'_n, post(c_n)))]$$

holds for every pair $(i_0, i_1)$, $(k_0, k_1)$, $(k'_0, k'_1)$, and $(c_0, c_1)$.

*Example 4.* In the system $\{r_1, r_2, r_3, r_4\}$ (Fig. 4), the registration and posting phases are pairwise independent. On the other hand, in the system $\{r_1, r_2, r_5, r_6, r_7, r_8\}$ or $\{r_1, r_2, r_9, r_{10}, r_{11}, r_{12}\}$, the registration and posting phases are not pairwise independent.

Cases 2 and 3 can be extended to show the sequential compositionality of role interchangeability. To obtain these results, we require the pairwise independence assumption.

CLAIM A.1. *Assume that the registration and posting phases are pairwise independent with respect to an observer $j$. Also suppose that every pair comprising an agent $k$ and an action $post(c)$ is role interchangeable with respect to $j$. Then, every pair comprising an agent $i$ and an action $submit(c)$ is role interchangeable as well.*

PROOF. Suppose that $(\mathcal{I}, r, m) \models \theta(i, submit(c))$ and $(\mathcal{I}, r, m) \models \theta(i', submit(c'))$. Then, there exist $k$ and $k'$ such that $(\mathcal{I}, r, m) \models \theta(i, use(k)) \wedge \theta(k, post(c))$ and $(\mathcal{I}, r, m) \models \theta(i', use(k')) \wedge \theta(k', post(c'))$. Because every pair comprising an agent $k$ and an action $post(c)$ is role interchangeable and because $(\mathcal{I}, r, m) \models \theta(k, post(c)) \wedge \theta(k', post(c'))$, we can say that $(\mathcal{I}, r, m) \models P_j[\theta(k', post(c)) \wedge \theta(k, post(c'))]$ holds. On the other hand, we have $(\mathcal{I}, r, m) \models \theta(i', use(k')) \wedge \theta(i, use(k))$. That is, $(\mathcal{I}, r, m) \models P_j[\theta(i', use(k')) \wedge \theta(i, use(k))]$ holds. So, by virtue of the pairwise independence assumption, $(\mathcal{I}, r, m) \models P_j[\theta(i', use(k')) \wedge \theta(k', post(c)) \wedge \theta(i, use(k)) \wedge \theta(k, post(c'))]$ holds. That is, $(\mathcal{I}, r, m) \models P_j[\theta(i', submit(c)) \wedge \theta(i, submit(c'))]$. This concludes the proof. □

CLAIM A.2. *Assume that the registration and posting phases are pairwise independent with respect to an observer $j$. Also suppose that every pair comprising an agent $i$ and an action $use(k)$ is role interchangeable with respect to $j$. Then, every pair comprising an agent $i$ and an action $submit(c)$ is role interchangeable as well.*

*Example 5.* In the system $\{r_1, r_2, r_5, r_6, r_7, r_8\}$ or $\{r_1, r_2, r_9, r_{10}, r_{11}, r_{12}\}$ (Fig. 4), every pair comprising an agent $k$ and an action $post(c)$ is role interchangeable as well as every pair comprising an agent $i$ and an action $use(k)$. However, the registration and posting phases are not pairwise independent. Consequently, in these systems, there exist some pairs comprising an agent $i$ and an action $submit(c)$ such that they are not role interchangeable.

Cases 8, 9, 10, and 11 in Table 1 are respectively derived from Cases 2, 3, 4, and 5 by replacing "up-to" anonymity/privacy properties with minimal anonymity/privacy properties. There are two problems in obtaining these derivations. First, consider Case 8 and its dual, Case 9, which are derived from Cases 2 and 3, respectively. Since the definition of minimal privacy/anonymity involves negative formulas, independence assumptions in positive-negative and negative-positive forms are helpful in these cases. Thus, we will use Lemmas A.2 and A.3 in Cases 8 and 9, respectively.

Second, consider Case 10 (which is derived from Case 4) and an intended example system consisting of the two indistinguishable runs $r_5$ and $r_6$ (Fig. 4). In $r_5$, $i_1$ uses $k_1$ and $k_2$ to post $c_1$ and $c_2$, respectively. In $r_6$, $i_1$ uses $k_1$ and $k_2$ to post $c_2$ and $c_1$, respectively. Thus, in the system $\{r_5, r_6\}$, every $use(k)$ performed by $i$ is maximally onymous and every $k$ performing $post(c)$ is minimally private, but $i$ performing $submit(c)$ is never minimally private. This is because although the posting actions performed by the pseudonyms $k_1$ and $k_2$ of $i_1$ are totally different, the submission actions performed by $i_1$ are defined using existential quantification over $k$ and thus both $\theta(i_1, submit(c_1))$ and $\theta(i_1, submit(c_2))$ hold in both $r_5$ and $r_6$. To avoid this, we assume that every real-name agent can be allowed to use at most one pseudonym in each run, that is, each $i$ is exclusive. This assumption will also be used in a generalization of Case 10, that is, Case 8. Note that to deal with Cases 9 and 11, we need a similar assumption that every possible article $c$ can be posted by at most one pseudonym $k$ in each run, that is, every $post(c)$ is exclusive, which is the "dual" of the assumption above.

CLAIM A.3. *Assume that the registration and posting phases are independent with respect to $j$. Suppose that the posting phase is exhaustive and that each $post(c)$ is exclusive as well as each $i$. Also suppose that every agent $k$ performing $post(c)$ is minimally private with respect to $j$. Then, every agent $i$ performing $submit(c)$ is minimally private.*

PROOF. Suppose that $(\mathcal{I}, r, m) \models \theta(i, submit(c))$. Then, there exists some $k$ in $I_P$ such that $(\mathcal{I}, r, m) \models \theta(i, use(k)) \wedge \theta(k, post(c))$. From $(\mathcal{I}, r, m) \models \theta(i, use(k))$, it is immediately seen that $(\mathcal{I}, r, m) \models P_j[\theta(i, use(k))]$. Because every $k$ performing $post(c)$ is minimally private and because $(\mathcal{I}, r, m) \models \theta(k, post(c))$, we can say that $(\mathcal{I}, r, m) \models P_j[\neg\theta(k, post(c))]$ holds. So, by virtue of Lemma A.2, $(\mathcal{I}, r, m) \models P_j[\theta(i, use(k)) \wedge \neg\theta(k, post(c))]$ holds. Since every real-name agent can be allowed to use at most one pseudonym in each run, this means that $(\mathcal{I}, r, m) \models P_j[\neg\theta(i, submit(c))]$ holds. □

CLAIM A.4. *Assume that the registration and posting phases are independent with respect to $j$. Suppose that the registration phase is exhaustive and that each $i$ is exclusive as well as each $post(c)$. Also suppose that every action $use(k)$ performed by $i$ is minimally anonymous with respect to $j$. Then, every action $submit(c)$ performed by $i$ is minimally anonymous.*

CLAIM A.5. *Suppose that the posting phase is exhaustive and that each $i$ is exclusive as well as each $post(c)$. Also suppose that every action $use(k)$ performed by $i$ is maximally onymous with respect to $j$. Moreover assume that every agent $k$ performing $post(c)$ is minimally private with respect to $j$. Then, every agent $i$ performing $submit(c)$ is minimally private.*

PROOF. This directly follows from Lemma 3.1 and Claim A.3. □

CLAIM A.6. *Suppose that the registration phase is exhaustive and that each $post(c)$ is exclusive as well as each $i$. Also suppose that every action $use(k)$ performed by $i$ is minimally anonymous with respect to $j$. In addition assume that every agent $k$ performing $post(c)$ is maximally identified with respect to $j$. Then, every action $submit(c)$ performed by $i$ is minimally anonymous.*

The final case shown in Table 1 indicates that if both the registration and posting phases guarantee linkability, then so does the entire system.

CLAIM A.7. *Suppose that every action $use(k)$ performed by $i$ is maximally onymous with respect to $j$ and that every agent $k$ performing $post(c)$ is maximally identified with respect to $j$. Then, every action $submit(c)$ performed by $i$ is maximally onymous.*

PROOF. Suppose that $(\mathcal{I}, r, m) \models \theta(i, submit(c))$. Then, there exists some $k$ in $I_P$ such that $(\mathcal{I}, r, m) \models \theta(i, use(k)) \wedge \theta(k, post(c))$. Because every action $use(k)$ performed by $i$ is maximally onymous and because every agent $k$ performing $post(c)$ is maximally identified, $(\mathcal{I}, r', m') \models \theta(i, use(k)) \wedge \theta(k, post(c))$ holds for every point $(r', m')$ such that $(r', m') \sim_j (r, m)$. This means that $(\mathcal{I}, r, m) \models K_j[\theta(i, submit(c))]$. □

## B. PARALLEL COMPOSITIONALITY: MORE CASES

In this appendix, we discuss Cases III to V shown in Table 1.

Cases III and IV are perfectly symmetric and deal with the parallel compositionality of minimal anonymity/privacy. Note that the independence assumption is unnecessary here.

CLAIM B.1. *Suppose that either $i$ performing $act_a(c)$ or $i$ performing $act_b(c)$ is minimally private with respect to $j$. Then, $i$ performing $act_p(c)$ is also minimally private.*

PROOF. Suppose that $(\mathcal{I}, r, m) \models \theta(i, act_p(c))$. Then, $(\mathcal{I}, r, m) \models \theta(i, act_a(c)) \wedge \theta(i, act_b(c))$ holds. Also assume that, say, $i$ performing $act_a(c)$ is minimally private. Then, based on the assumption of minimal privacy, $(\mathcal{I}, r, m) \models P_j[\neg\theta(i, act_a(c))]$ holds. This immediately implies that $(\mathcal{I}, r, m) \models P_j[\neg\theta(i, act_a(c)) \vee \neg\theta(i, act_b(c))]$ holds. That is, $(\mathcal{I}, r, m) \models P_j[\neg\theta(i, act_p(c))]$ holds. □

Case V in Table 2 indicates a trivial result on the parallel compositionality of linkability.

CLAIM B.2. *Suppose that both $i$ performing $act_a(c)$ and $i$ performing $act_b(c)$ are maximally identified with respect to $j$. Then, $i$ performing $act_p(c)$ is also maximally identified.*

## C. INDEPENDENCE-AS-ANONYMITY/PRIVACY INTERPRETATION

In this appendix, we discuss that the independence assumption shown in Sect. 3 could be viewed by itself as a "meta-level" abstraction of the anonymity or privacy property.

We first introduce two additional conditions regarding our anonymous members-only bulletin board system. We say that the bulletin board system satisfies *backward causality* provided that if $k$ posts $c$, then there exists some $i$ such that $i$ uses $k$. This is formulated as $\mathcal{I} \models \theta(k, post(c)) \Rightarrow \bigvee_{i \in I_R} \theta(i, use(k))$. Backward causality can be regarded as a natural assumption in that every posted article should be related by some real-name agent; however, it is not a mandatory assumption because in some cases, certain auxiliary pseudonyms may post some dummy articles to enhance the privacy of real-name agents. We may also assume *forward causality*, which means that if $i$ uses $k$, then there exists some $c$ such that $k$ posts $c$.

It is immediately seen that the definition of independence is equivalent to stating that $\mathcal{I} \models \theta(k', post(c)) \Rightarrow \bigwedge_{i,k}(P_j[\theta(i, use(k))] \Rightarrow P_j[\theta(i, use(k)) \wedge \theta(k', post(c))])$ holds for every $k'$ and $c$. If we assume backward causality, then this is also equivalent to that for every $i'$, $k'$, and $c$,

$$\mathcal{I} \models \theta(i', use(k')) \wedge \theta(k', post(c)) \Rightarrow$$
$$\bigwedge_{i,k}(P_j[\theta(i, use(k))] \Rightarrow P_j[\theta(i, use(k)) \wedge \theta(k', post(c))])$$

holds. If we abuse the notation and write $\Theta(\theta(i, use(k)), coexist(\theta(k', post(c))))$ for $\theta(i, use(k)) \wedge \theta(k', post(c))$, which means a "meta-level" link between "first-class" links $\theta(i, use(k))$ and $\theta(k', post(c))$, then the above equivalent transformation indicates that the independence assumption can be viewed as a certain, abstract form of "anonymity." More specifically, the obtained, equivalent formula means that an "action" $coexist(\theta(k', post(c)))$ performed by an "agent" $\theta(i', use(k'))$ is anonymous up to a certain "anonymity set" with respect to $j$. Alternatively, if we assume forward causality, the independence assumption can be viewed as an abstract form of "privacy." When we apply our framework to the compositional verification of the anonymity or privacy property of a specific example, it will often be a key task to show that the independence assumption holds. The above remark suggests a possibility that we can use conventional proof methods for anonymity/privacy when showing the independence assumption, although we do not go into detail here.